# Observation of nonvolatile magneto-thermal switching in superconductors

H. Arima[1], Md. R. Kasem[1], H. Sepehri-Amin[2], F. Ando[2], K. Uchida[2], Y. Kinoshita[3],

M. Tokunaga[3], Y. Mizuguchi[1]*

[1]*Department of Physics, Tokyo Metropolitan University; Hachioji, 192-0397, Japan.*

[2]*National Institute for Materials Science; Tsukuba, 305-0047, Japan.*

[3] *Institute for Solid State Physics, University of Tokyo, Kashiwa, 277-8581, Japan*

*Corresponding author: mizugu@tmu.ac.jp

**Applying a magnetic field to a solid changes its thermal-transport properties. Although such magneto-thermal-transport phenomena are usually small effects, giant magneto-thermal resistance has recently been observed in spintronic materials[1,2] and superconductors[3,4], opening up new possibilities in thermal management technologies. However, the thermal conductivity conventionally changes only when a magnetic field is applied due to the absence of nonvolatility, which limits potential applications of thermal switching devices[5,6]. Here, we report the observation of nonvolatile thermal switching that changes the thermal conductivity when a magnetic field is applied and retains the value even when the**

**field is turned off. This unconventional magneto-thermal switching, surprisingly, arises in commercial Sn-Pb solders and is realized by phase-separated superconducting states and resultant nonuniform magnetic flux distributions. This result confirms the versatility of the observed phenomenon and aids the development of active solid-state thermal management devices.**

Thermal switching is a growing and crucial component of thermal management[5,6] because heat flow control is essential to achieve high efficiencies in electronic devices. In particular, thermal switching without mechanical motion is important to control the heat flow in solids; metal-insulator transition[7], electrochemical interactions[8], electric fields[9], and magnetic fields[10] ($H$) have been used to switch the thermal conductivity ($\kappa$) of materials. Magneto-thermal switching (MTS) is a promising technology because a huge MTS has been observed in spintronic multilayer films[1,2] and superconducting materials[3,4]. After investigations on various MTS materials, the MTS ratio (MTSR), which is defined as $[\kappa(H) - \kappa(0)] / \kappa(0)$, now exceeds 1000% without observation of *nonvolatile* characteristics of MTS. If nonvolatile MTS materials whose $\kappa(H)$ value can be maintained at zero fields after experiencing $H$ are obtained, they can provide a new pathway to achieve efficient thermal management in solids. In this study, we show that conventional (commercial) Sn-Pb solders exhibit nonvolatile MTSR of 150%, which is defined as $[\kappa(0, \text{demagnetized}) - \kappa(0, \text{initial})] / \kappa(0, \text{initial})$.

MTS of superconductors is achieved below its superconducting transition temperature ($T_c$) by forming Cooper pairs in the superconducting state, where the Cooper pairs do not transfer heat, which results in the reduction of carrier $\kappa$. The MTSR of superconductors can be extremely large; MTSR > 1000% has been confirmed in highly pure Pb[10]. Although the working temperature of superconductors is quite low, they are potentially suitable for the thermal management of low-temperature electronic devices[11,12]. However, pure superconductors do not exhibit nonvolatile characteristics in the $H$ dependence of $\kappa$. In this study, we investigate the MTS characteristics of Sn-Pb solders and observe that simple solders exhibit nonvolatile MTS. We conclude that the mechanism of nonvolatile MTS in the solders is based on trapped magnetic flux, as discussed later. Flux trapping in Sn-Pb solders was investigated through magnetization measurements several decades ago[13,14], in which the Sn-Pb solders were simply regarded as type-II superconductors. However, the Sn-Pb solders are actually composite (phase-separated) materials composed of two type-I superconductors with different $T_c$, i.e., Sn ($T_c$ = 3.7 K) and Pb ($T_c$ = 7.2 K). Here, we propose that such composite superconductors trap magnetic flux nonuniformly and give rise to nonvolatile MTS.

Here, we briefly introduce the magnetic flux trapping in superconductors. Superconductors are mainly categorized into type-I and type-II, based on the difference in their reaction to applied $H$[15]. In type-I superconductors, perfect diamagnetism is observed up to the critical field ($H_c$), and the superconducting states are quickly suppressed by applying further fields. Therefore, magnetic flux is expelled from ideal type-I superconductors in their superconducting state. Type-II superconductors have two critical fields: a lower critical field ($H_{c1}$) and an upper critical field ($H_{c2}$). At $H < H_{c1}$, perfect diamagnetism is also observed in type-II superconductors, but at $H_{c1} < H < H_{c2}$, the magnetic flux can coexist with the superconducting states. The magnetic flux inside the type-II superconductors is quantized, where the vortices with a normal-conducting core (and a lattice of vortices) are formed[15]. The vortices have been detected experimentally and investigated by various techniques[16–19]. Further, thin type-I superconductor films also exhibit vortex states when the film thickness is very thin or the films contain weak pinning centers[20,21]. Another phenomenon of trapped magnetic flux was observed in a superconductor hollow cylinder or ring[22]. Because of the shielding supercurrents in the superconducting cylinder, fluxes are trapped inside the cylinder. This phenomenon occurs in both type-I and type-II superconductor cylinders. As the present sample is a composite superconductor composed of type-I Sn and Pb, both the above mechanisms of flux trapping can emerge. The size of the grains in the phase-separated Sn-Pb solder is of the μm scale, as shown later, which is close to the vortex size of sub-μm to several μm. Therefore, the Sn grains can host vortices and lose their bulk superconducting nature due to the formation of normal-conducting cores even though Sn is a type-I superconductor. Furthermore, magnetic fluxes may be trapped in the Sn regions by shielding currents in the Pb regions. In addition, the Sn regions start conducting normally (non-superconducting) because of trapped field of $H > H_c$ (Sn). Although the mechanisms behind the flux trapping are known, unexpectedly strong flux trapping in a bulk composite composed of type-I superconductors with different $T_c$ would provide new insights into functionalities and applications of superconductors.

**Nonvolatile magneto-thermal switching in Sn-Pb solder**

The most important result of this work is the observation of nonvolatile MTS in Sn-Pb solders. It is widely known that solders are phase-separated composites, but the utilization of unique superconducting states emerging in the phase-separated solders has not attracted much attention. Here, we show the nonvolatile characteristics of MTS at $T$

= 2.5, 3.0, and 4.2 K as examples. The schematic images of the concept of nonvolatile MTS in solders are shown in Fig. 1. At the initial state (Fig. 1a), the whole sample is superconductive, and the $\kappa$ is low due to the suppression of carrier heat transfer. At $H > H_c$, the superconducting states of the solder are totally suppressed (Fig. 1b), and $\kappa$ is increased by the revival of thermal conduction by charge carriers. The nonvolatility of MTS is observed by reducing $H$ after experiencing a large $H$. As shown in Fig. 1c, high $\kappa$ is retained even after removing external fields (at $H = 0$ Oe), which is achieved by the magnetic fluxes trapped in the Sn regions. The nonvolatility of $\kappa$ implies that several Sn grains lose the bulk nature of superconductivity and are close to normal-conducting states because of the trapped magnetic fluxes. The trapping of a large number of magnetic fluxes in the Sn regions of Sn-Pb solders had not been a common understanding in the field of pure and applied science of superconductors.

We measured the temperature and field dependences of $\kappa$ for commercial Sn45-Pb55 solders using a four-probe method (Fig. 2a). Figure 2b shows the temperature dependences of $\kappa$ measured at $H = 0$ Oe after zero-field cooling (ZFC) and field cooling under $H = 1500$ Oe (FC). In addition, the FC data measured at $H = 1500$ Oe are plotted together with data measured at $H = 0$ Oe (after ZFC and FC). The difference in the $\kappa$-$T$ curve appears below 7 K, which is due to the emergence of superconductivity in the solder (at $T_c$ for Pb; see magnetization data shown in Fig. 3a). As shown in Fig. 2b, we find that the ZFC and FC data exhibit clear differences when these measurements are performed after removing the applied magnetic field ($H = 0$ Oe) in the measurement system. At $H = 1500$ Oe (FC), the decrease in $\kappa$ at low temperatures was totally suppressed because the superconducting states of the solder were destroyed. Because the FC ($H = 0$ Oe) data exhibited an intermediate trend between ZFC ($H = 0$ Oe) and FC ($H = 1500$ Oe), it is clear that magnetic fluxes, less than 1500 Oe, were trapped in the solder sample after the FC under 1500 Oe.

Figures 2c–2f display the $\kappa$-$H$ curve measured at $T$ = 2.5, 3.0, 4.2, and 8.0 K. Here, error bars are not displayed for clarity, but the data with error bars are displayed in Extended Data Figure 1. No MTS was observed at $T$ = 8.0 K because the temperature was higher than $T_c$ of the solder. At $T$ = 2.5 K, a clear MTS was observed in the initial increments of $H$ from 0 to 1700 Oe. Further, by decreasing $H$ from 1700 to -1700 Oe, $\kappa$ slightly decreases but does not reach the initial value of $\kappa$ at $H$ = 0 Oe. At around -800 to -1000 Oe, an anomaly is seen, which is related to the critical field (Extended Data Figure 2). By increasing $H$ from -1700 to 1700 Oe, a similar anomaly was observed between 800 to 1000 Oe, but the value of $\kappa$ never returned to the initial value. The $\kappa$-$H$ data clearly shows the nonvolatile MTS characteristic in the solder. The nonvolatile MTSR was about 150%, as shown in Fig. 2c. At $T$ = 3.0 and 4.2 K, similar nonvolatile MTS trends were observed, while the MTSR decreased with increasing temperature. One of the reasons why nonvolatile MTS was observed at $T$ = 4.2 K (> $T_c$ of Sn) would be explained by the partial suppression of the superconducting states of the Pb regions by the trapped fluxes. Another reason would be weak superconducting states in the Sn regions achieved by the proximity effects in the initial state at $T$ = 4.2 K. After field experience, Meissner states cannot be achieved due to the presence of trapped fluxes. Comparable MTS characteristics were obtained for an Sn45-Pb55 solder wire in an as-purchased form with a φ1.6 mm cross-section (Extended Data Figure 3). Furthermore, we examined the MTS on a flux-cored solder with a different composition (Sn60-Pb40) and observed similar nonvolatile MTS (Extended Data Figure 4). Therefore, nonvolatile MTS is a common behavior in various Sn-Pb solders.

**Characterization of superconducting properties and phase separation of Sn-Pb solder**

To understand the causes of nonvolatile MTS in solders, the superconducting properties were investigated by measuring the magnetization ($M$) and specific heat ($C$). Figures 3a and 3b show the $T$ dependence of $M$ measured at approximately 10 Oe after ZFC and FC (under 1500 Oe); ZFC data exhibits diamagnetism below 7.2 K, but FC data exhibits ferromagnetic-like signals below 7.2 K. Similar significant differences in the $M$-$T$ between ZFC and FC have been observed in type-II superconductors; for a recent observation example, superhydrides (hydrogen-rich superconductors) exhibit similar ferromagnetic-like $M$-$T$ behavior after FC[23]. This behavior is explained by the trapped flux in the type-II superconductors. In contrast, our Sn-Pb solder sample was composed of type-I Sn and Pb, which is clearly different from the former case. As shown in Fig. 4, the elemental mapping analysis revealed that there are phase-separated Sn and Pb regions with a typical size of 5-20 μm. In the μm-scale order, the superconducting states of Pb can penetrate the Sn region, which causes the single-step superconducting transition shown in Fig. 3a. Instead, the FC data in Fig. 3b exhibits a ferromagnetic-like behavior with a transition temperature of 7.2 K, which is the $T_c$ of Pb. This suggests that magnetic fluxes were trapped in the solder at temperatures below $T_c$ of Pb. Therefore, the material would have characteristics similar to that observed in type-II superconductors after FC. As a fact, we observed the broadening of the temperature dependence of resistivity under magnetic fields (Extended Data Figure 2). The trend is commonly observed in type-II superconductors with strong pinning.[24] The fact suggests that the trapped fluxes are thermally fluctuating, which is consistent with the result shown in Fig. 3b.

To further characterize the magnetic properties, the $H$ dependence of magnetization ($4\pi M$) was measured at $T$ = 2.5, 3.0, and 4.2 K (Fig. 3c and 3d), where the data was corrected by a demagnetization factor. With decreasing temperature, the size of the $4\pi M$-$H$ hysteresis becomes larger, which suggests the enhancement of a critical current density ($J_c$) and critical field. However, we noticed there was no large change between $T$ = 3.0 and 4.2 K. As $T_c$ of pure Sn is 3.7 K, the absence of a large change in the $4\pi M$-$H$ hysteresis at around 3.7 K indicates that the characteristics of the emerging superconducting currents are governed by Pb in the solder. To further understand what is happening in the solder under magnetic fields, we plotted inner magnetic flux density ($B$), which is given by $B = H + 4\pi M$, in Fig. 3e. When the solder was zero-field-cooled to $T$ = 2.5 K, the initial $B$-$H$ curve exhibited perfect diamagnetism (Meissner states) up to about 500 Oe. Then, $B$ becomes equal to $H$, which indicates the suppression of the superconducting states at $H > 700$ Oe. When decreasing the field from $H > 1000$ Oe, an anomaly appears at $H \sim 700$ Oe, where the superconducting states of Pb emerge, and magnetic fluxes are trapped inside the solder. Even at $H = 0$ Oe, the $B$ remains a large value of ~500 G, which is consistent with a previous work[14]. Based on those facts, we concluded that magnetic fluxes with $B \sim 500$ G can be trapped after FC or application of $H$ greater than 700 Oe in the solder, and the flux trapping in the Sn regions is the origin of nonvolatile MTS.

To prove the assumption above, we measured the temperature dependence of zero-field specific heat ($C$) after ZFC and FC (under 1500 Oe). The results, including the data measured under a magnetic field ($H$ = 1500 Oe), are summarized in Extended Data Figure 5. In Fig. 3f, the $C$ data in a form of $C/T$ after removing normal states values, estimated by subtracting the data at 1500 Oe, are plotted as a function of temperature. As shown in Extended Data Figure 5, from the analysis of low-temperature $C$ at $H = 1500$ Oe using the low-temperature approximation of $C = \gamma T + \beta T^3 + \delta T^5$, Debye temperature ($\theta_D$) and electronic specific heat coefficient ($\gamma$) were estimated as 128.8(4) K and 2.14(5) mJ/K$^2$mol, respectively. As $\theta_D$ for Pb and Sn are about 105 and 199 K[25], respectively, the obtained $\theta_D$ for the Sn45-Pb55 solder would be reasonable. For both ZFC and FC data at $H = 0$ Oe, jumps at $T_c$ of Pb were observed; the large value of the specific heat jump $\Delta C/\gamma T_c$ ~1.4 for the Pb regions (42% in molar ratio to Sn) is clearly greater than the value expected by weak-coupling BCS model[26] ($\Delta C/\gamma T_c = 1.43$), which is consistent with the strong-coupling nature of Pb and alloyed Pb[27]. Noticeably, the jump at $T_c$ of Sn ($T$ ~ 3.7 K) corresponding to the emergence of the superconducting states of Sn was observed only for the ZFC data, suggesting that the Sn regions do not undergo a *bulk* superconducting transition with a large entropy change after FC. Therefore, the trapped fluxes should be mainly present in the Sn regions. The Sn-Pb phase-separation size (Fig. 4) may be a situation suitable for stabile flux trapping and suppression of bulk superconductivity of the Sn regions in solders. We did not observe a clear decrease in the amount of FC magnetization until after at least two days, as shown in Extended Data Figure 6, which is evidencing strong trapping of fluxes and a merit when using this phenomenon in applications.

**Discussion**

To explore the possibility of initializing (ON-to-OFF switching) $\kappa$ using magnetic field control, we focused on the characteristic point of the *B-H* loop, as shown in Fig. 3e. Coming back from positive high $H$, $B$ crosses the $H = 0$ Oe line with finite positive $B$. Then, $B$ reaches zero at -470 Oe. Therefore, we investigated whether $B$ can return to the origin $(H, B) = (0$ Oe, $0$ G$)$. Figure 5a shows the *M-H* loop when measuring $M$ at $H = 0 \rightarrow 1500 \rightarrow 0 \rightarrow -470 \rightarrow 0$ Oe; orange arrows in Figs. 5a and 5b explain this field experience process. $4\pi M$ reaches the origin position of the loop, and the inner magnetic flux density $B$ also reaches its origin, as shown in Fig. 5b. These results indicate that certain magnetic-field controls can return net $B$ to the initial value. We expected a reduction in $\kappa$ due to the recovery of superconductivity of the Sn regions using the same magnetic field control, but, as shown in Figs. 5c–5e, $\kappa$ does not reach the initial value. Those results imply the absence of bulk superconductivity in the Sn regions even in the state of net $B = 0$ G achieved after the process of $H = 0 \rightarrow 1500 \rightarrow 0 \rightarrow -470 \rightarrow 0$ Oe. From the results on net $B$ and $\kappa$, we concluded that local $B$ does not become zero, where the compensation of fluxes parallel to $+H$ and $-H$ resulted in net $B = 0$ G. Although we have not directly measured the direction of trapped fluxes (and/or vortices) of the solder, the coexistence of fluxes with opposite directions can be assumed from the results. As magnetic field control cannot achieve initialization of $\kappa$ in the solders, other methods should be developed to achieve nonvolatile MTS with initialization functionality. Heating up to $T > T_c$ ($T > 7.2$ K for the solder) or flowing current greater than the critical current density will work to reset the flux-trapping states and initialize the $\kappa$ value, because of breaking the superconducting states. In Fig. 5f, the temperature evolution of $\kappa$ measured at $H = 0$ Oe after FC (1500 Oe) is shown. The initialization of $\kappa$ by increasing the sample temperature to $T > T_c$ is achieved.

To further obtain experimental proofs for the flux trapping in the Sn regions, we performed magneto-optical (MO) imaging for the Sn45-Pb55 solder at $T = 2.5$ K. The obtained images are shown in Extended Data Figure 7. Image (i) corresponds to the initial state after ZFC where no magnetic flux is trapped. When the magnetic field is $H = 1500$ Oe, greater than critical field of the solder, μm-order structures are observed in image (ii). These structures indicate the uniform presence of magnetic fluxes in the normal conducting states. After decreasing $H$ to 0 Oe, magnetic fluxes are expelled from the Pb regions and trapped in the Sn regions only. In image (iii), we observe blurriness of the structures and the emergence of contrast different from image (ii). By applying negative $H$, reversed trends are seen. In image (iii), the light parts would be Pb-rich regions, and the dark parts would be Sn-rich regions. Because of the bulk-sensitive mechanism of MO imaging, we just observed the blurriness and the changes in contrast because the Sn regions are inhomogeneously distributed along the thickness direction. If magnetic fluxes are trapped at the grain boundaries, the MO image should show a uniform image in the flux-trapping states. Therefore, the present results can exclude the scenario of trapping at the grain boundaries. To obtain further evidence of the magnetic fluxes in the Sn regions, surface-sensitive techniques are needed to directly observe trapped fluxes in the solders.

For future application of this phenomenon, tunability of nonvolatile MTS is preferred. Here, we investigated the Sn-content dependence of nonvolatile MTS (Extended Data Figure 8). For Sn90-Pb10, clear nonvolatility is not observed, but for Sn10-Pb90, nonvolatile MTS of ~ 300% is observed. In addition, the $\kappa$ values change with changing Sn amount. Furthermore, we evaluated minor-loop characteristics of $\kappa$-$H$ for the Sn45-Pb55 solder (Extended Data Figure 9). As shown in Extended Data Figure 10i, nonvolatile MTS is determined by the maximum field. These tunability of nonvolatile MTS by composition and magnetic field provide new thermal management functionalities.

Here, we demonstrated that Sn-Pb solders exhibit nonvolatile MTS by utilizing the superconducting states of Pb and flux trapping in the Sn regions. In the solders, the coexistence of two superconducting phases with different $T_c$, $T_c = 7.2$ K for Pb and $T_c = 3.7$ K for Sn, gives rise to nonvolatile MTS. In addition, the trapped fluxes large enough to suppress bulk superconductivity of the Sn regions are essential for this phenomenon; hence, $B > H_c$ for Sn was the preferred condition in the present case. The concept that superconductor composites can work as nonvolatile MTS materials is quite simple and applicable to various pairs of superconductors. For example, using high-$T_c$ superconductors in the nonvolatile MTS composite would increase the working temperature of the nonvolatile MTS phenomena by making composite with metals, alloys, or intermetallic compounds. Optimizing phase-separation conditions should enhance the switching ratio and flexibility of nonvolatile MTS.

Furthermore, since the Sn-Pb solder is widely used in electrical wiring, the large nonvolatile thermal switching and magnetic flux trapping in the solders should significantly affect low-temperature transport measurement techniques that were believed to be already established. Therefore, understanding the magneto-transport phenomena in solders is essential for reliable transport measurements at low temperatures under magnetic fields.

**References**

1. Kimling, J. et al. Spin-dependent thermal transport perpendicular to the planes of Co/Cu multilayers. *Phys. Rev. B* **91**, 144405 (2015).

2. Nakayama, H. et al. Above-room-temperature giant thermal conductivity switching in spintronic multilayer. *Appl. Phys. Lett.* **118**, 042409 (2021).

3. Yoshida, M., Kasem, Md. R., Yamashita, A., Uchida, K. & Mizuguchi, Y. Magneto-thermal-switching properties of superconducting Nb. *Appl. Phys. Express* **16**, 033002 (2023).

4. Yoshida, M., Arima, H., Yamashita, A., Uchida, K. & Mizuguchi, Y. Large magneto-thermal-switching ratio in superconducting Pb wires. arXiv:2305.08332.

5. Li, N. et al. Phononics: Manipulating heat flow with electronic analogs and beyond. *Rev. Mod. Phys.* **84**, 1045 (2012).

6. Wehmeyer, G., Yabuki, T., Monachon, C., Wu, J. & Dames, C. Thermal diodes, regulators, and switches: Physical mechanisms and potential applications. *Appl. Phys. Rev.* **4**, 041304 (2017).

7. Nishimura, Y. et al. Electronic and Lattice Thermal Conductivity Switching by 3D−2D Crystal Structure Transition in Nonequilibrium $(Pb_{1-x}Sn_x)Se$. *Adv. Electron. Mater.* **8**, 2200024 (2022).

8. Cho, J. et al. Electrochemically tunable thermal conductivity of lithium cobalt oxide. *Nat. Commun.* **5**, 4035 (2014).

9. Ihlefeld, J. F. et al. Room-temperature voltage tunable phonon thermal conductivity via reconfigurable interfaces in ferroelectric thin films. *Nano Lett.* **15**, 1791 (2015).

10. Kimling, J., Gooth, J. & Nielsch, K. Anisotropic magnetothermal resistance in Ni nanowires. *Phys. Rev. B* **87**, 094409 (2013).

11. Huang, H. L., Wu, D., Fan, D. & Zhu, X. Superconducting quantum computing: a review. *Sci. China Info. Sci.* **63**, 180501 (2020).

12. Kjaergaard, M. et al. Superconducting Qubits: Current State of Play. *Ann. Rev. Condens. Matter Phys.* **11**, 369–395 (2020).

13. Livingston, J. D. Magnetic Properties of Superconducting Lead-Base Alloys. *Phys. Rev.* **129**, 1943–1949 (1963).

14. Furuya, S., Tominaga, A. & Narahara, Y. The Wall-Thickness Dependence of Magnetic Shielding or Trapping in a Low-Field Superconductor, $Pb_{40}Sn_{60}$. *J. Low Temp. Phys.*, **53**, 477–485 (1983).

15. Abrikosov, A. A. Nobel Lecture: Type-II superconductors and the vortex lattice. *Rev. Mod. Phys.* **76**, 975–979 (2004).

16. Harada, K. et al. Real-time observation of vortex lattices in a superconductor by electron microscopy. *Nature* **360**, 51–53 (1992).

17. Hess, H. F., Robinson, R. B. & Waszczak, J. V. STM spectroscopy of vortex cores and the flux lattice. *Physica B* **169**, 422–431 (1991).

18. Wells, F. S., Pan, A. V., Wang, X. R., Fedoseev, S. A. & Hilgenkamp, H. Analysis of low-field isotropic vortex glass containing vortex groups in $YBa_2Cu_3O_{7-x}$ thin films visualized by scanning SQUID microscopy. *Sci. Rep.* **5**, 8677 (2015).

19. Iguchi, Y. et al. Superconducting vortices carrying a temperature-dependent fraction of the flux quantum. *Science* **380**, 1244–1247 (2023).

20. Dolan, G. J. & Silcox, J. Critical Thicknesses in Superconducting Thin Films. *Phys. Rev. Lett.* **30**, 603–606 (1973).

21. Ge, J., Gutierrez, J., Cuppens, J. & Moshchalkov, V. V. Observation of single flux quantum vortices in the intermediate state of a type-I superconducting film. *Phys. Rev. B* **88**, 174503 (2013).

22. Doll, R. & Näbauer, M. Experimental Proof of Magnetic Flux Quantization in a Superconducting Ring. *Phys. Rev. Lett.* **7**, 51–52 (1961).

23. Minkov, V. S., Ksenofontov, V., Bud'ko, S. L., Talantsev, E. F. & Eremets, M. I. Magnetic flux trapping in hydrogen-rich high-temperature superconductors. *Nat. Phys.* (2023): https://doi.org/10.1038/s41567-023-02089-1.

24. Xing, X. et al. Anisotropic Ginzburg–Landau scaling of $H_{c2}$ and transport properties of 112-type $Ca_{0.8}La_{0.2}Fe_{0.98}Co_{0.02}As_2$ single crystal. *Supercond. Sci. Technol.* **29**, 055005 (2016).

25. Stewart, G. R. Measurement of low-temperature specific heat. *Rev. Sci. Instrum.* **54**, 1–11 (1983).

26. Bardeen, J., Cooper, L. N. & Schrieffer, J. R. Theory of Superconductivity. *Phys. Rev.* **108**, 1175–1204 (1957).

27. Padamsee, H., Neighbor, J. E. & Shiffman, C. A. Quasiparticle Phenomenology for Thermodynamics of Strong-Coupling Superconductors. *J. Low Temp. Phys.* **12**, 387–411 (1973).

28. Kinoshita, Y., Miyakawa, T., Xu, X. & Tokunaga, M. Long-distance polarizing microscope system combined with solenoid-type magnet for microscopy and simultaneous measurement of physical parameters. *Rev. Sci. Instrum.* **93**, 073702 (2022).

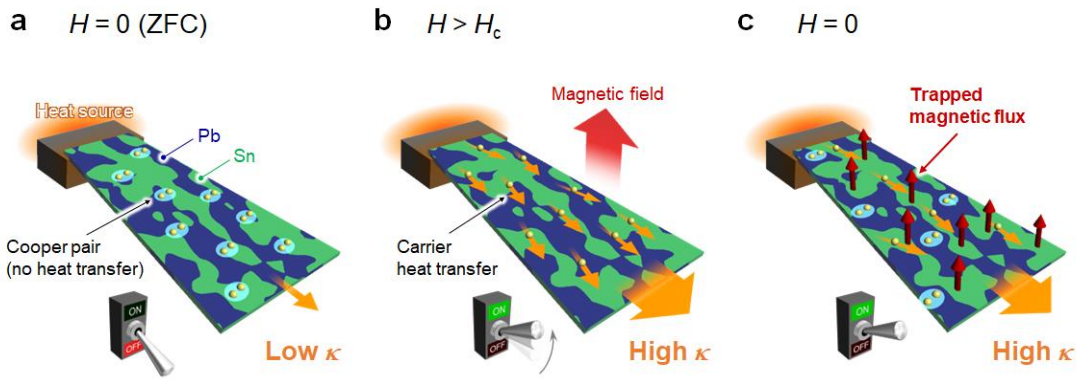

**Figure 1 | Schematic images of nonvolatile magneto-thermal switching observed in commercial solder (Sn45-Pb55). a,** Initial state with low thermal conductivity ($\kappa$) after zero-field cooling (ZFC). The schematic image of a switch (OFF) denotes the low-$\kappa$ state. **b,** State under a magnetic field ($H$) higher than the critical field ($H_c$). Magnetic field lines can penetrate whole samples because both Pb and Sn are in normal conducting states. In this state, $\kappa$ is high (ON). **c,** State at $H = 0$ after experiencing $H > H_c$. Pb is in the superconducting state, and the magnetic field lines do not penetrate the Pb regions. Fluxes trapped in the Sn regions cannot be released even at $H = 0$, which results in the suppression of bulk superconductivity in the Sn regions. In this state, nonvolatile MTS with high $\kappa$ (ON) can be observed.

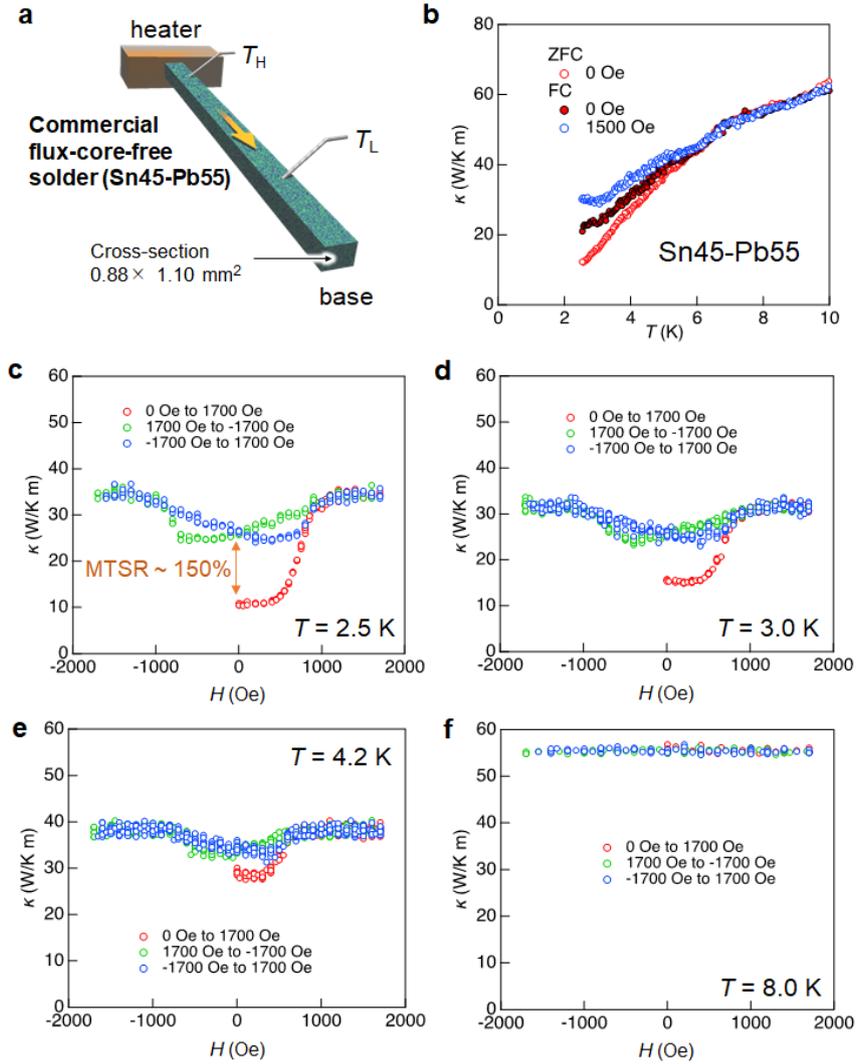

**Figure 2 | Nonvolatile magneto-thermal switching characteristics of the flux-core-free solder (Sn45-Pb55). a,** Schematic image of the measured sample with a cross-sectional area of 0.88×1.10 mm². $T_H$ and $T_L$ denote two thermometers. The purchased solder with a diameter of 1.6 mm was polished into a uniform rectangular bar. **b,** Temperature ($T$) dependence of $\kappa$ measured at $H = 0$ Oe after ZFC and field cooling at $H = 1500$ Oe. FC (0 Oe) means that the sample was field-cooled from 10 K to 2.5 K, and the data were taken after reducing the external field. In addition, we show the data measured at $H = 1500$ Oe (FC 1500 Oe). **c–f,** $\kappa$-$H$ curves measured at $T = 2.5, 3.0, 4.2,$ and 8.0 K.

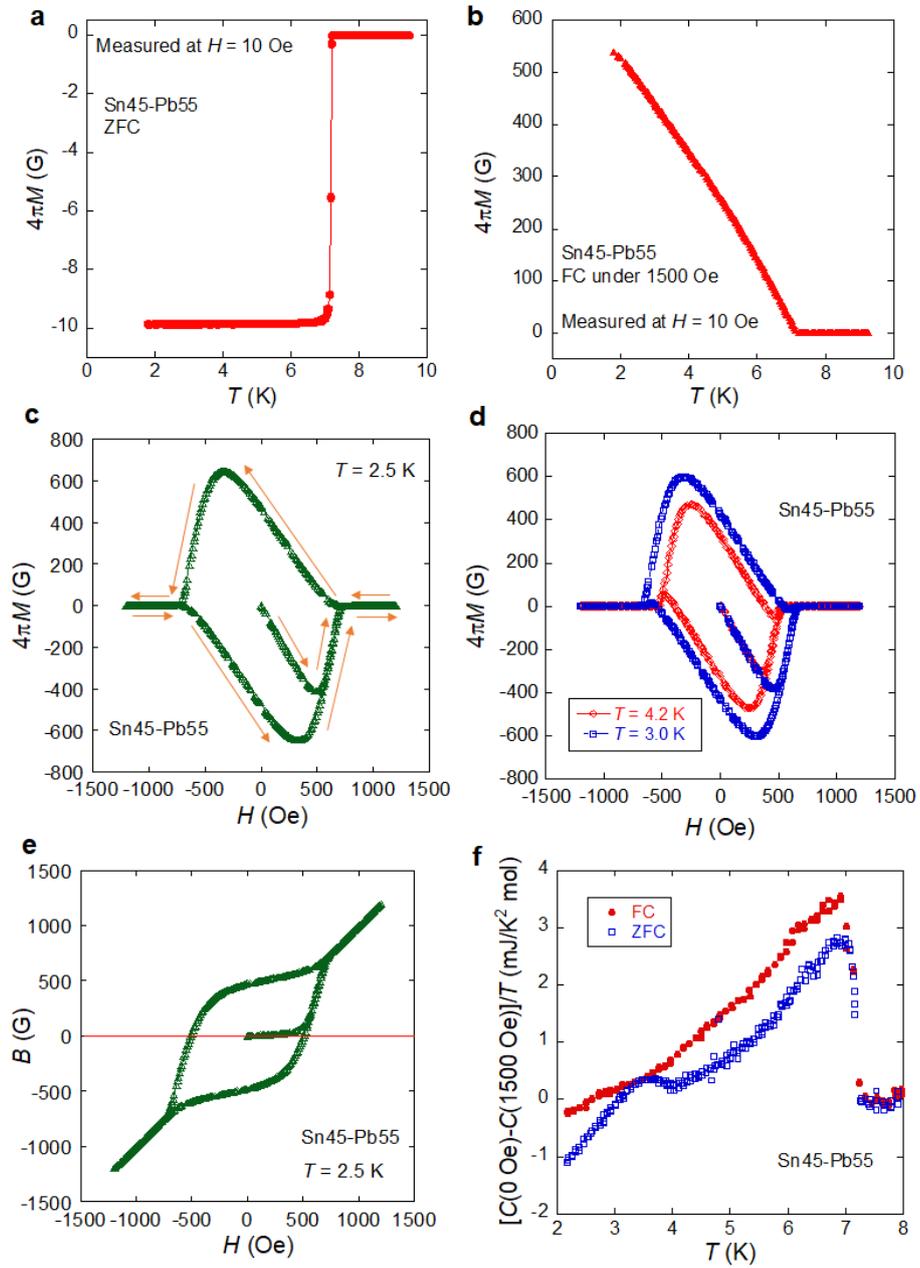

**Figure 3 | Superconducting properties of the solder Sn45-Pb55. a,b,** $T$ dependence of magnetization ($4\pi M$) measured at about 10 Oe after zero-field cooling (ZFC) and field cooling under 1500 Oe (FC). **c, d,** $M$-$H$ curves measured at $T$ = 2.5, 3.0, and 4.2 K. **e,** $H$ dependence of inner magnetic flux density ($B$). **f,** $T$ dependence of residual specific heat

estimated by $C$ (0 Oe) – $C$ (1500 Oe) in a form of $C/T$. The superconducting transition of Sn ($T_c$ = 3.7 K) is seen in the ZFC data only.

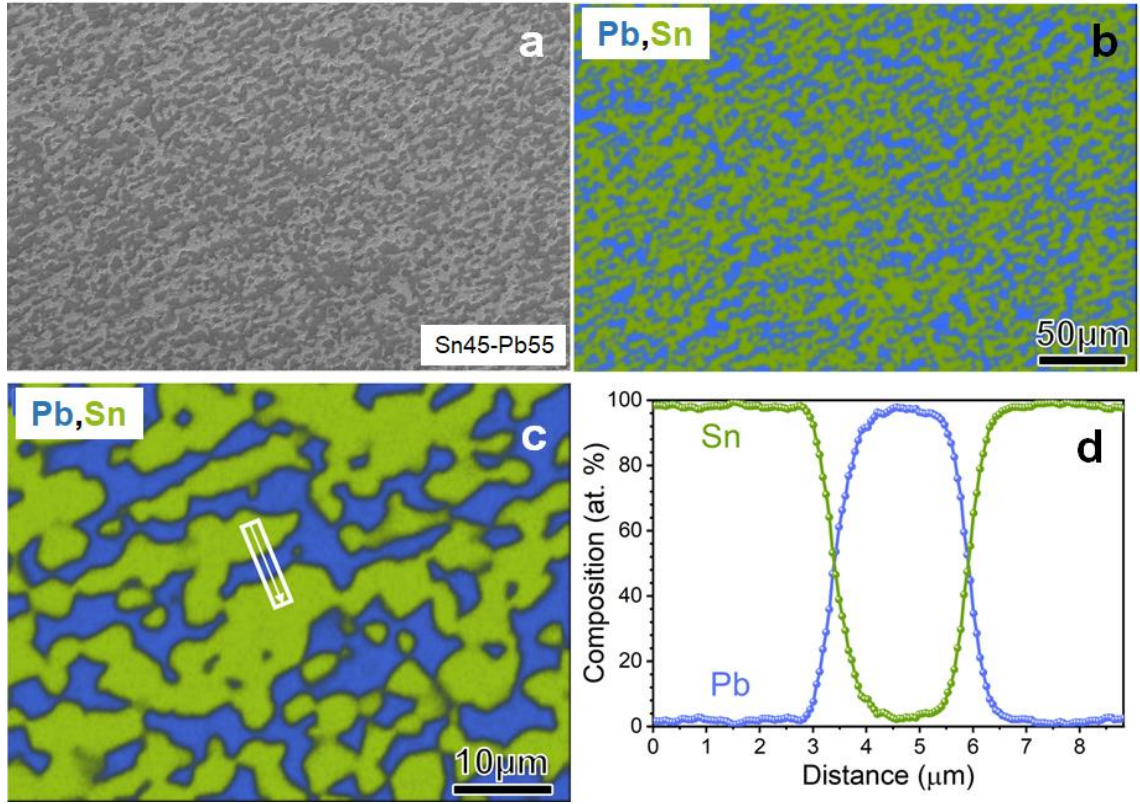

**Figure 4 | µm-scale phase separation of Pb and Sn in the solder. a,** Scanning-electron microscope (SEM) image on the polished surface of the solder (Sn45-Pb55). **b,c,** Elemental mapping by energy-dispersive X-ray spectroscopy (EDX). **d,** Line profiles of the compositions of Sn and Pb along the white arrow in **c**.

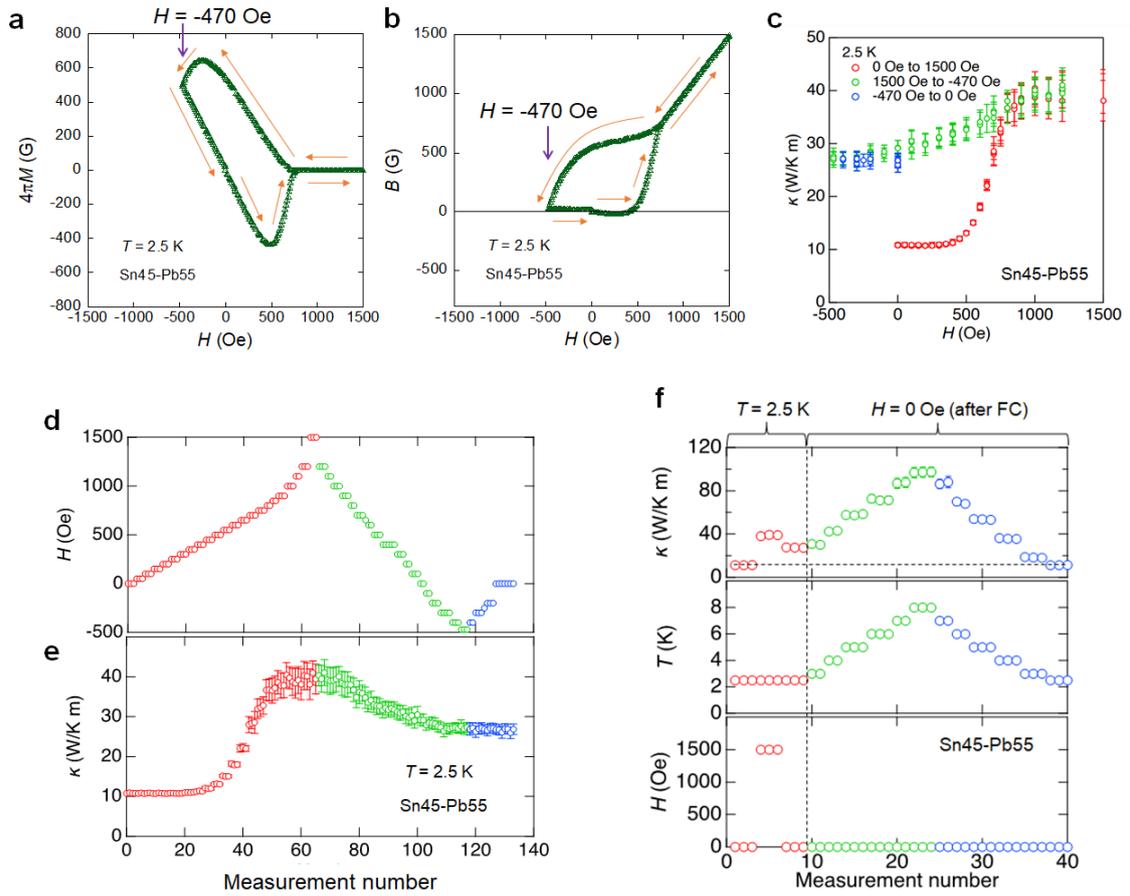

**Figure 5 | Minor loop measurements and initialization of $\kappa$. a,b,** $H$ dependence of $4\pi M$ and $B$ at $T = 2.5$ K for the solder (Sn45-Pb55), measured when $H$ was swept between 1500 Oe and -470 Oe. See the arrows for the guide for measurement order. **c,** $H$ dependence of $\kappa$, measured when $H$ was swept from 0 to 1500 Oe (red), from 1500 to -470 Oe (green), and from -470 to 0 Oe (blue). It is clearly shown that the initial $\kappa$ cannot be recovered by any magnetic-field control. **d,e,** Measurement number dependence of $H$ and $\kappa$ at $T = 2.5$ K, extracted from the data in **c**. **f,** Initialization of $\kappa$ by heating the sample above $T_c$. Measurement number dependence of $\kappa$, $T$ and $H$. First, the ON state was produced by applying $H = 1500$ Oe and reducing the field to zero at $T = 2.5$ K (measurement number: 1–9). For number 10–24, temperature was gradually increased to

$T = 8.0$ K. Then, temperature decreased to $T = 2.5$ K for number 25–40, and the initial $\kappa$ is recovered as indicated by the dashed line.

**Methods**

**Samples.**   We used commercial solders: flux-core-less Sn45-Pb55 solder wires (φ1.6 mm, TAIYO ELECTRIC IND. CO., LTD.) and flux-cored Sn60-Pb40 solder wires (φ0.8 mm, HOZAN). The data shown in the main text was taken on a polished sample of Sn45-Pb55 solder wire. The purity of the Sn45-Pb55 solder wire was investigated by X-ray Fluorescence (XRF), and the actual Sn ratio was confirmed as 43.84(2)% in the weight ratio and $Sn_{0.58}Pb_{0.42}$ in the molar ratio. The Cu impurity with a weight ratio of 0.2% was detected by XRF. The Sn10-Pb90 and Sn90-Pb10 solders with 99.9% purity were purchased from SASAKI SOLDER INDUSTRY CO., LTD.

**Characterization.**   Scanning-electron microscope (SEM) and energy-dispersive X-ray spectroscopy (EDX) were used to analyze chemical compositions on the surface of the solders. The images for Sn45-Pb55 shown in the main text were taken using Carl Zeiss Cross-Beam 1540ESB and that for Sn60-Pb40 was taken using TM3030 (Hitachi Hightech). XRF was performed using JSX-1000S (JEOL).

**Physical property measurements.**   Thermal conductivity ($\kappa$) was measured by means of Physical Property Measurement System (PPMS, Quantum Design) with a thermal transport option (TTO) using a four-probe steady-state method with heater, two thermometers, and base-temperature terminal. The lengths between two thermometers attached to the measured samples were 55.5 mm for the Sn45-Pb55 rectangular bar with a cross-section area of 0.88×1.10 mm$^2$ (reported in the main text), 65.0 mm for the Sn45-Pb55 wire with a φ1.6 mm in diameter (reported in Extended Data Figure 3), and 44.5 mm for Sn60-Pb40 (reported in Extended Data Figure 4). Due to the limitation of the sample-room space of the TTO stage, the sample was screwed to store inside with four probes, a heater, two thermometers, and thermal base. Typical measurement duration for a single measurement was 30 seconds. The main result ($\kappa$-$H$ at 2.5 K) was measured manually (not in a sequence mode) to check the temperature stability and the reliability of the relaxation curves.

Magnetization was measured by a superconducting quantum interference device (SQUID) magnetometry on Magnetic Property Measurement System (MPMS3, Quantum Design) with a VSM mode. Specific heat was measured on PPMS by a relaxation mode. The sample was attached on a stage using APIEZON N grease. Electrical resistivity was measured on PPMS by a four-probe method under magnetic fields.

**Magneto-optical imaging**

For magneto-optical imaging, we used PPMS and an infinity-corrected objective lens inserted into the sample space by the microscope which is set above the PPMS at approximately 1 m from the sample position[28]. The images shown here were prepared by subtracting the images taken at $T = 8.0$ K (normal state) from those taken at $T = 2.5$ K and normalized by data taken at $T = 8.0$ K.

**Acknowledgments**

We thank O. Miura, A. Yamashita, M. Yoshida, T. D. Matsuda, K. Hattori, R. Kurita, Y. Oikawa, H. Fujihisa, A. Kikkawa, and T. Machida for supports in experiments and fruitful discussion on the results. This work was partly supported by JST-ERATO (JPMJER2201), TMU Research Project for Emergent Future Society, the joint research in the Institute for Solid State Physics, the University of Tokyo (202306-HMBXX-0090), and Tokyo Government Advanced Research (H31-1).## Author contributions

K.U. and Y.M. planned and supervised the study. H.A., H.S.A., K.U., Y.K., M.T., and Y.M. designed the experiments. H.A., M.R.K., H.S.A., Y.K., M.T., and Y.M. collected and analyzed the data. H.A., F.A., K.U., and Y.M. prepared the manuscript. All the authors discussed the results, developed the explanation of the experiments, and commented on the manuscript.

## Competing interest declaration

The authors declare no competing financial interests.

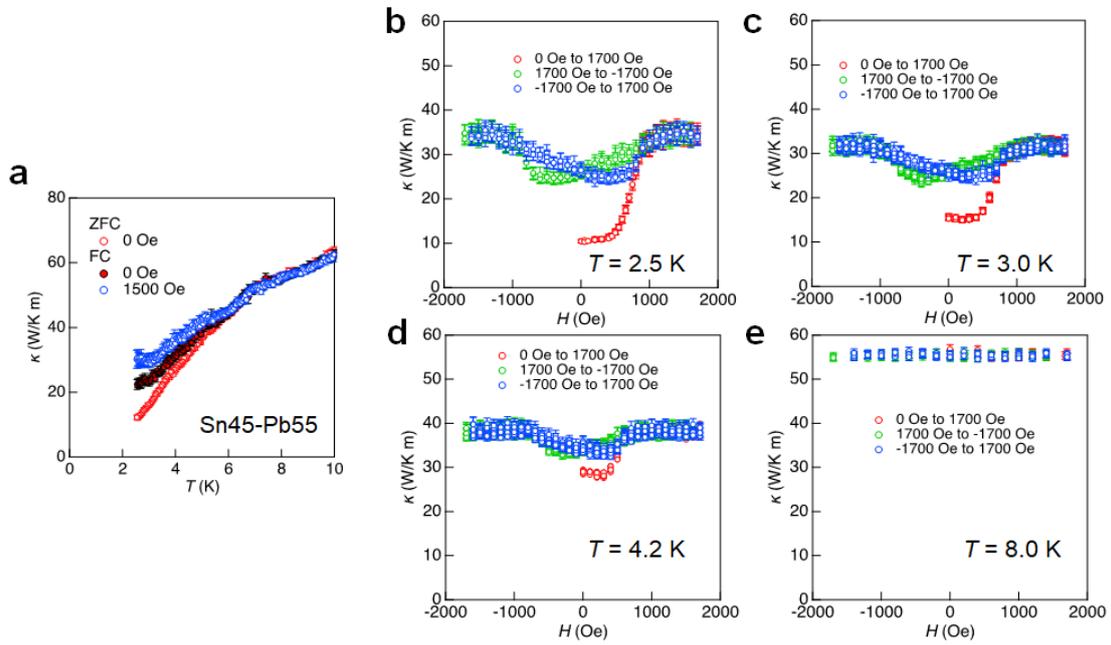

**Extended Data Figure 1 | Measurement errors on nonvolatile magneto-thermal switching characteristics of the flux-core-free solder (Sn45-Pb55). a**, $\kappa$-$T$ plot for the flux-core-free Sn45-Pb55 solder (rectangular-shaped sample shown in the main text) with error bars (standard deviations). FC (0 Oe) means that the sample was field-cooled from 10 K to 2.5 K, and the data were taken after reducing the external field. **b–e,** $\kappa$-$T$ plots taken at $T$ = 2.5, 3.0, 4.2, and 8.0 K with error bars.

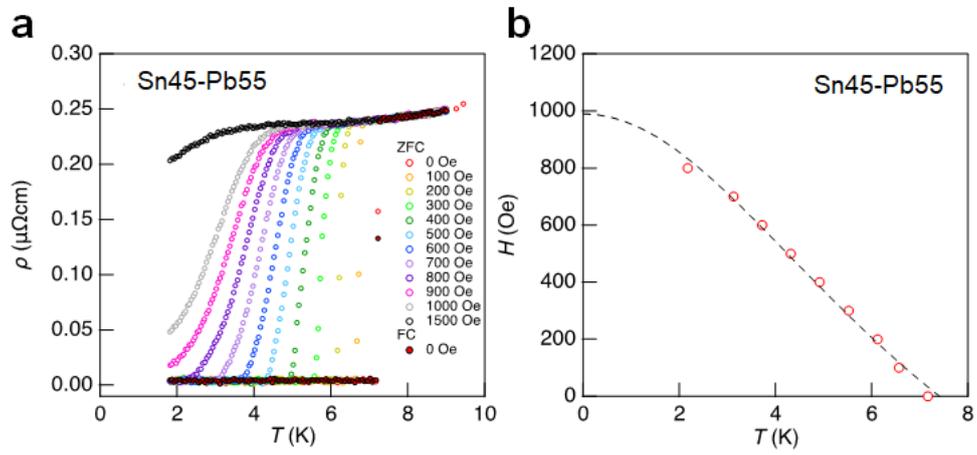

**Extended Data Figure 2. Transport properties of the flux-core-free solder (Sn45-Pb55).** **a,** Temperature dependence of the electrical resistivity for the flux-core-free Sn45-Pb55 solder. **b,** $H$-$T$ phase diagram estimated from the temperature dependence data.

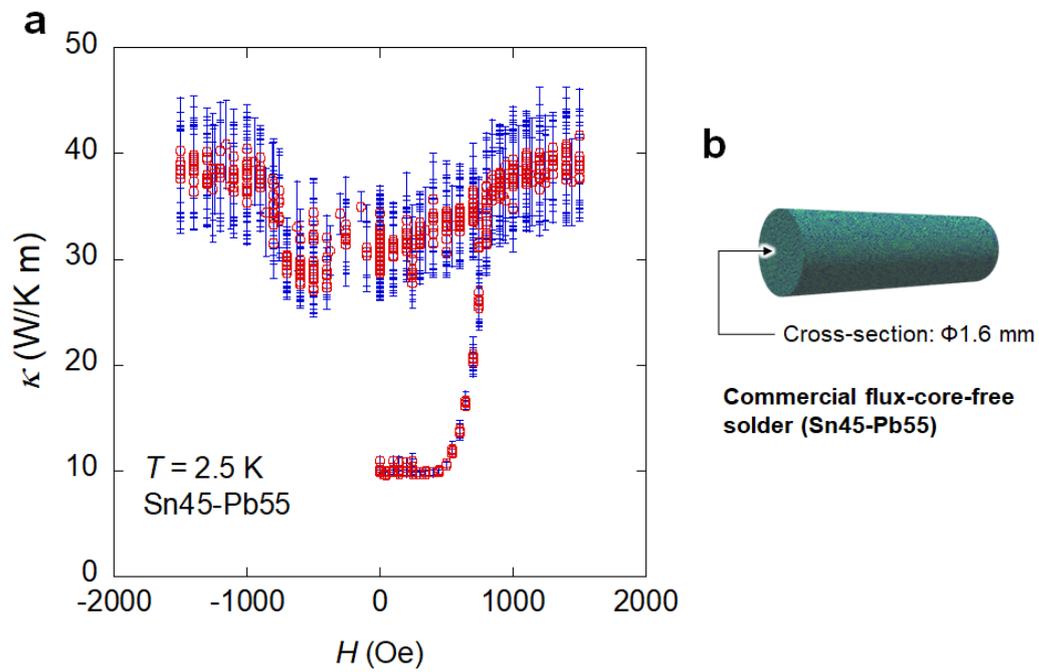

**Extended Data Figure 3 | Nonvolatile MTS for wire-shaped Sn45-Pb55. a,** $\kappa$-$H$ plot for a φ1.6 mm wire of the flux-core-free Sn45-Pb55 sample (without polishing) taken at $T$ = 2.5 K. **b,** Schematic image of the measured solder sample.

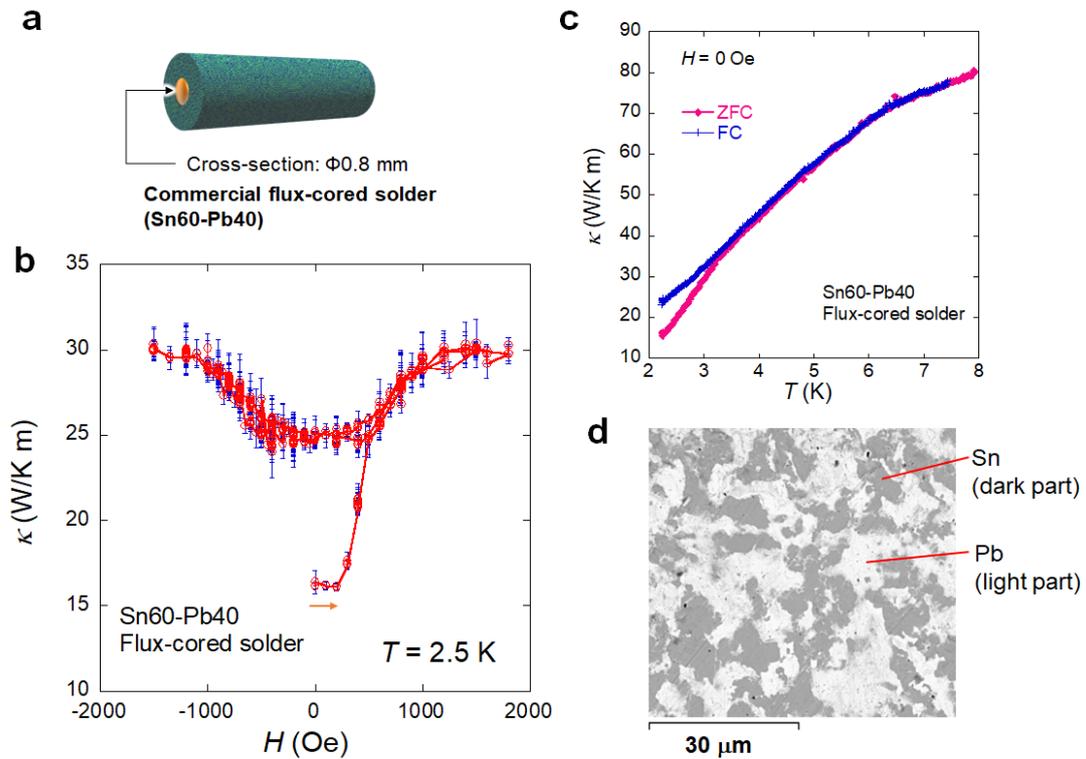

**Extended Data Figure 4 | Nonvolatile MTS for the flux-cored solder. a,** Schematic image of the measured solder sample. **b,** $\kappa$-$H$ plot for a $\varphi 0.8$ mm wire of the flux-cored Sn60-Pb40 sample (without polishing) taken at $T = 2.5$ K. **c,** $T$ dependence of $\kappa$ measured at $H = 0$ Oe after ZFC and FC (FC under $H = 1500$ Oe). **d,** SEM back-scattering image.

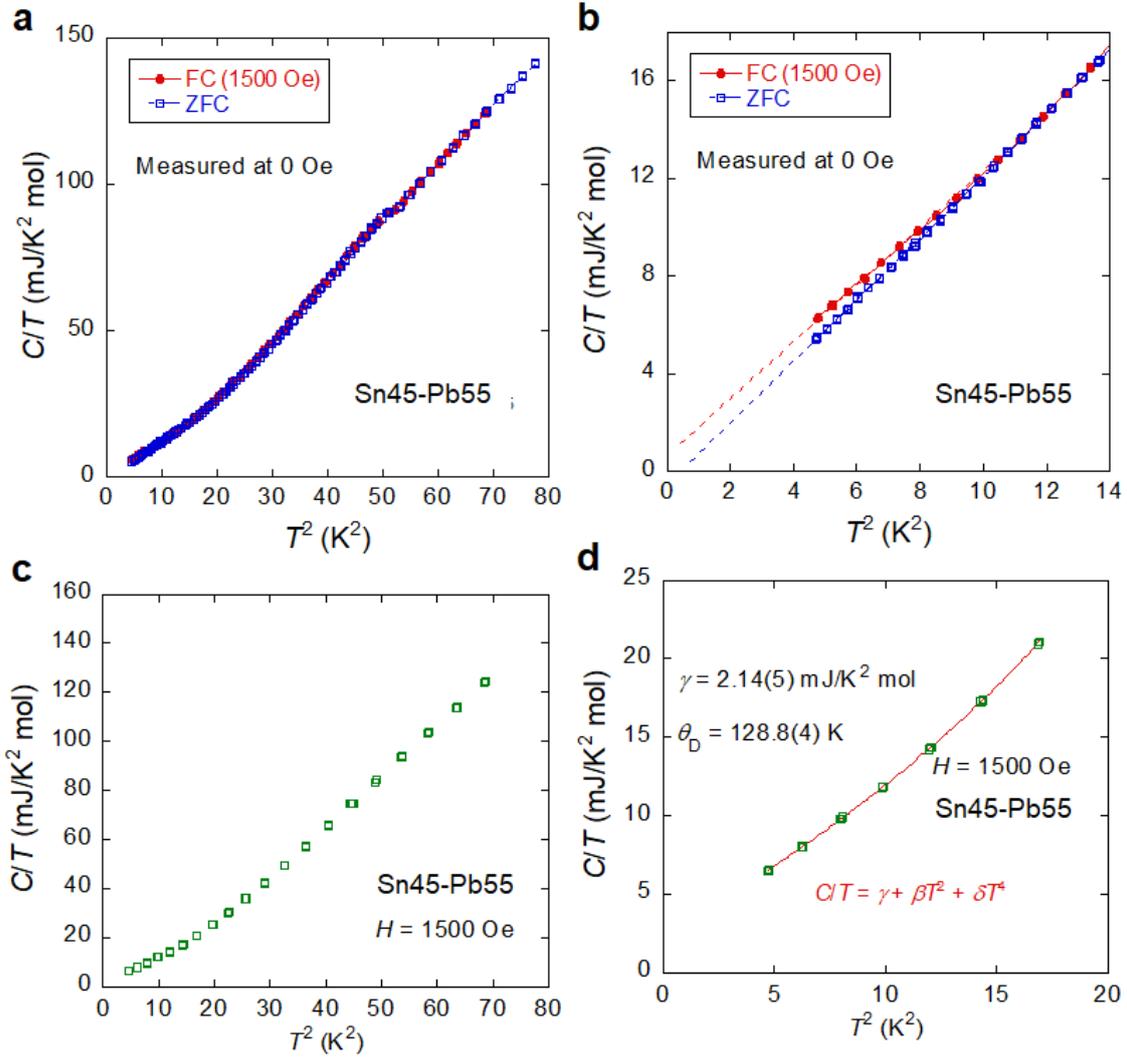

**Extended Data Figure 5 | Specific heat data for the flux-core-free solder (Sn45-Pb55).**
**a,b,** Squared temperature ($T^2$) dependence of the specific heat in a form of $C/T$ measured at $H = 0$ Oe for the flux-core-free Sn45-Pb55 solder. The data were taken after zero-field cooling (ZFC) and field cooling at $H = 1500$ Oe (FC). In **Extended Data Figure 1b**, clear differences appear below transition temperature ($T_c$) of Sn. **c,** $T^2$ dependence of $C/T$ measured at $H = 1500$ Oe for Sn45-Pb55. **d,** Low-temperature analysis for the data $H = 1500$ Oe where $C/T$ was fitted using a formula of $C/T = \gamma + \beta T^2 + \delta T^4$. Debye temperature ($\theta_D$) was calculated as 128.8(4) K using $\beta$ and the relation of $\beta = 12\pi^4 N k_B/\theta_D{}^3$ where $N$ and $k_B$ are the number of atoms and Boltzmann constant, respectively.

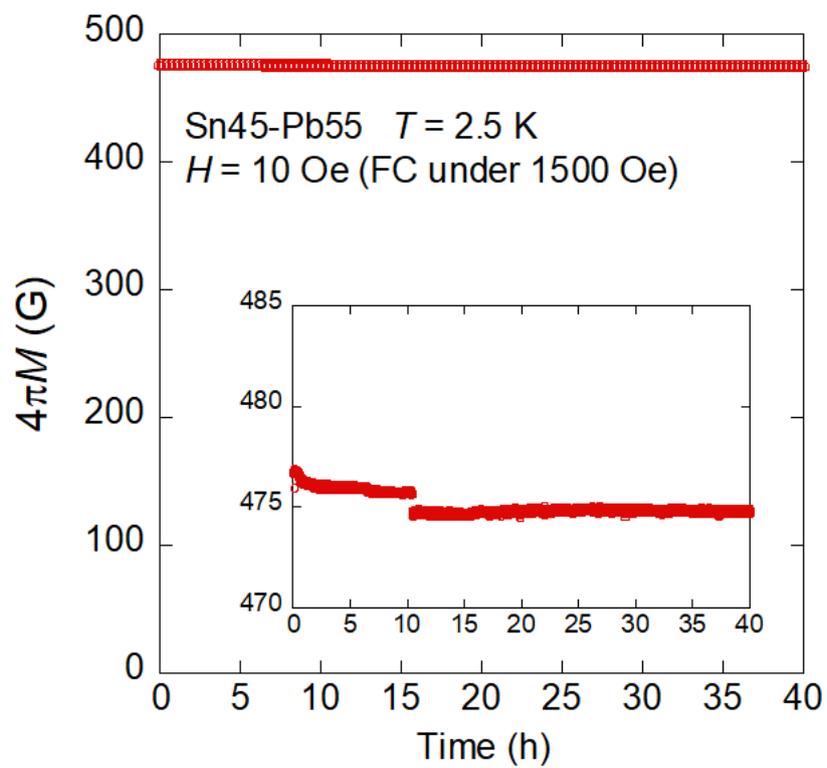

**Extended Data Figure 6 | Stability of the trapped flux.** Time dependence of magnetization for flux-core-free Sn45-Pb55 solder measured at $T = 2.5$ K for 40 h.

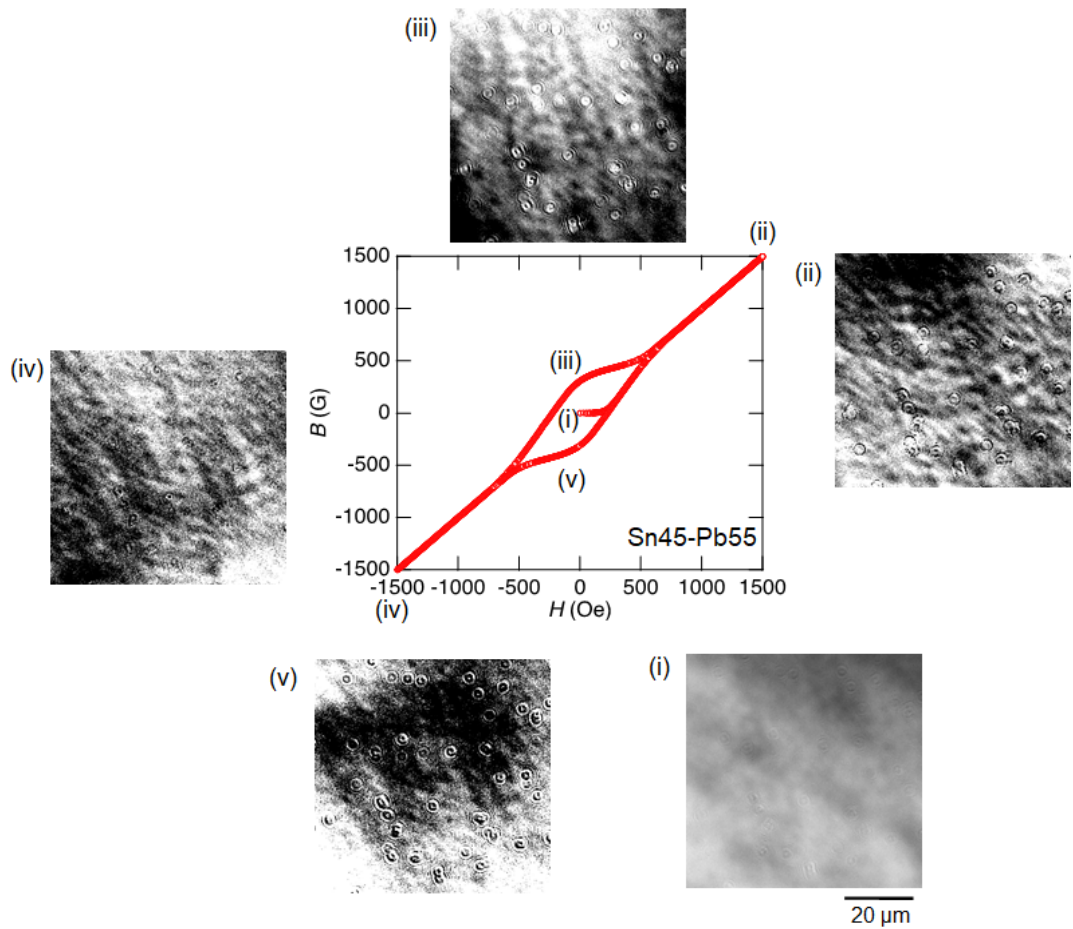

**Extended Data Figure 7 | Magneto-optical (MO) imaging.** We performed MO imaging on the Sn45-Pb55 solder at $T = 2.5$ K after different magnetic-field experiences. The bubble-like particles on all images are extrinsic impurities on the microscope system. Light parts correspond to positive magnetic fluxes. Image (i) is taken after ZFC, and no magnetic fluxes are observed. Image (ii) and (iv) are taken at $H = 1500$ and $-1500$ Oe, respectively; µm-order structures, which indicate the uniform presence of magnetic fluxes in the normal conducting states, are observed. Image (iii) and (v) are taken at $H = 0$ Oe but with positive and negative flux trapping, respectively. Here, similar structures observed in image (ii) and (iv) are present, but they are blurred. The blurriness would be caused by the presence of inhomogeneity of Sn/Pb regions along the thickness direction because of the sample thickness of about 100 µm. The noticeable features are the appearance of different contrast between (ii) and (iii) and between (iv) and (v). The dark parts in image (iii) and light parts in image (v) correspond the Pb-rich regions in the Meissner states.

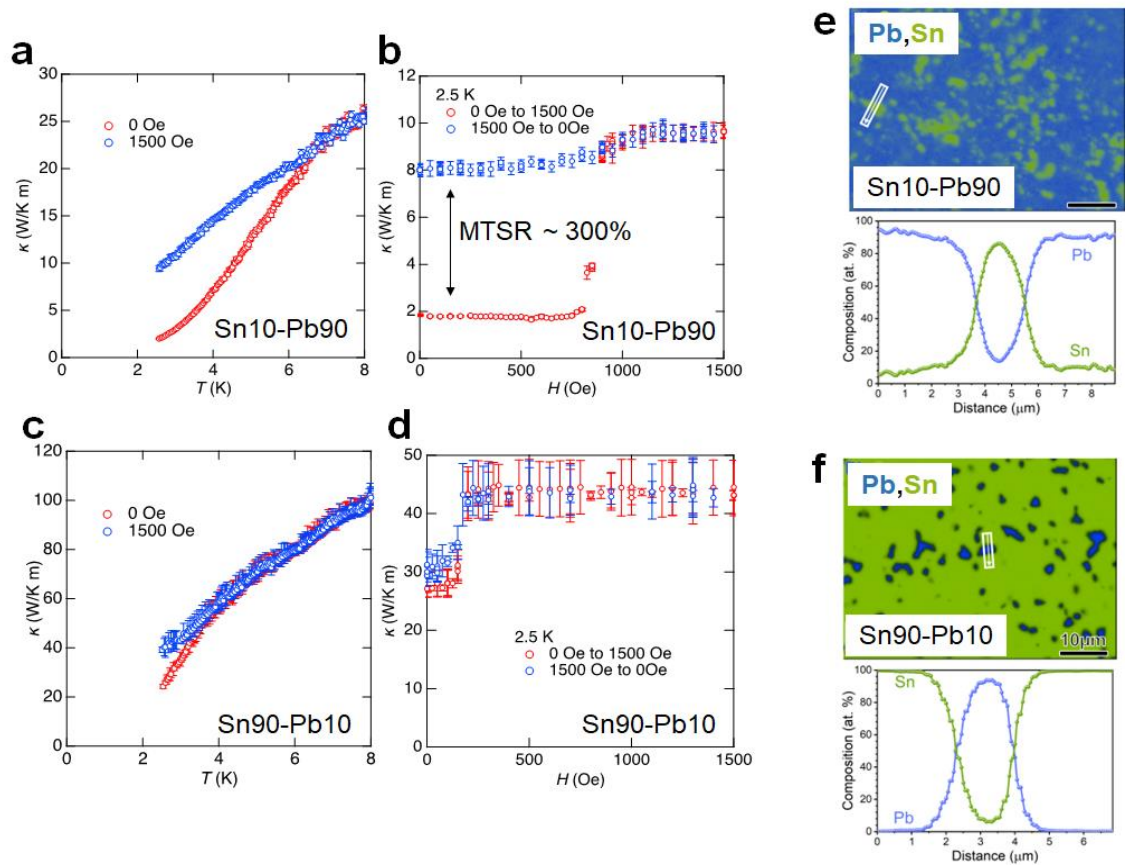

**Extended Data Figure 8 | Tunable nonvolatile MTSR by Sn-Pb composition. a,b**, Temperature dependence and field dependence of $\kappa$ for the Sn10-Pb90 solder wire. **c,d**, Temperature dependence and field dependence of $\kappa$ for the Sn90-Pb10 solder wire. **e,f**, SEM-EDX analyses results for the Sn10-Pb90 and Sn90-Pb10 solders.

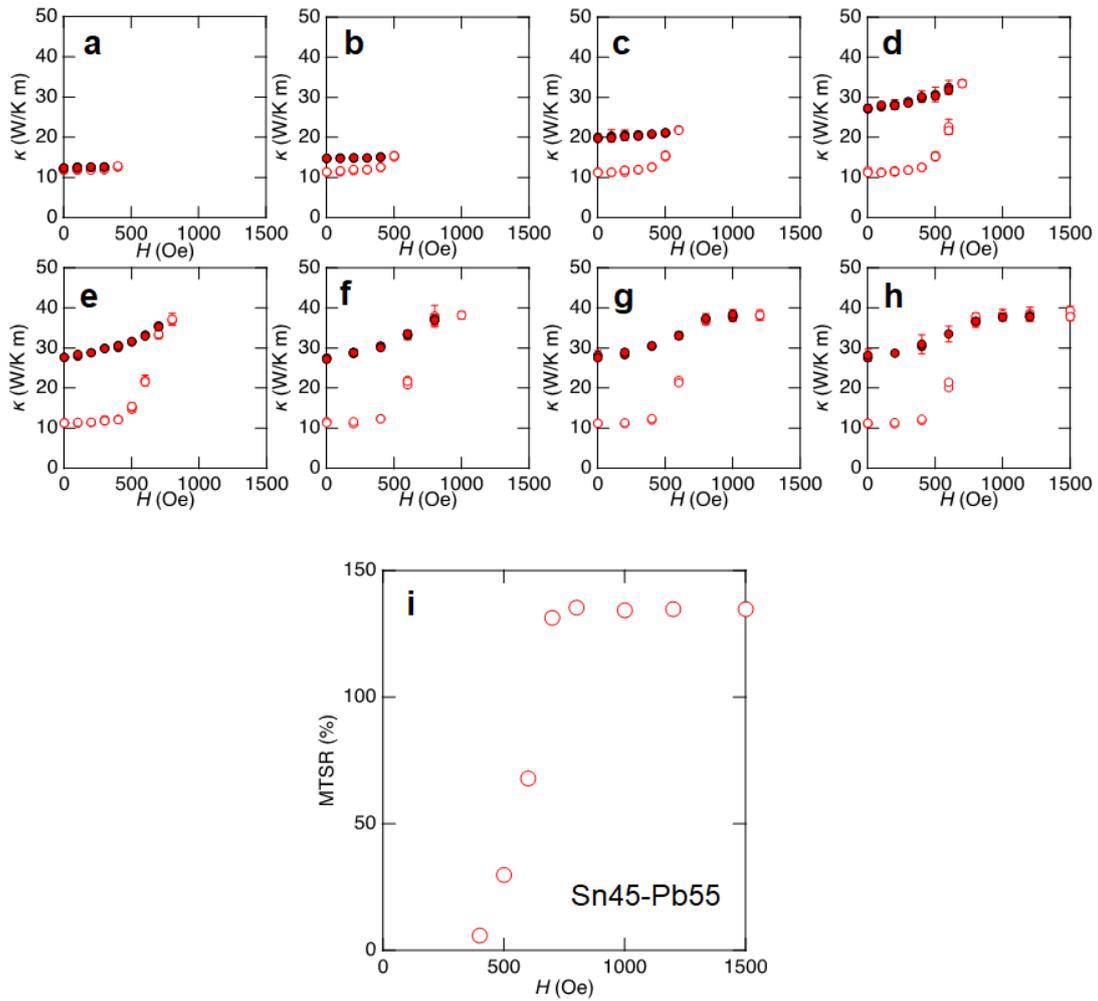

**Extended Data Figure 9 | Tunability of the ON value by tuning applied magnetic field. a–h,** Field dependence of $\kappa$ for the Sn45-Pb55 solder with different minor loops. **i,** Maximum field dependence of MTSR.

33